\documentclass[twocolumn]{aastex631}
\usepackage{xspace}
\usepackage{amsmath}
\usepackage{physics}
\usepackage{amsfonts} 
\usepackage{soul, hyperref}
\usepackage{lipsum, babel, ulem, soul, cancel}
\usepackage[shortlabels]{enumitem}

\newcommand{\orate}{occurrence rate\xspace}

\newcommand{\gyro}{gyrochronology\xspace}
\newcommand{\deteff}{detection efficiency\xspace}
\newcommand{\iso}{isochrone\xspace}

\newcommand{\resp}{respectively\xspace}
\newcommand{\Pdet}{$P_\mathrm{det}$\xspace}
\newcommand{\Pgeom}{$P_\mathrm{geom}$\xspace}
\newcommand{\Pwin}{$P_\mathrm{win}$\xspace}
\newcommand{\Porb}{$P_\mathrm{orb}$\xspace}
\newcommand{\solmass}{M$_\odot$\xspace}
\newcommand{\solrad}{R$_\odot$\xspace}
\newcommand{\earthrad}{R$_\oplus$\xspace}
\newcommand{\app}{$\sim$}

\newcommand{\teff}{T$_\textrm{eff}$\xspace}
\newcommand{\rp}{$Rp$\xspace}

\newcommand{\pval}{$p-$value\xspace}
\newcommand{\pvals}{$p-$values\xspace}

\newcommand{\kepler}{\textit{Kepler}\xspace}
\newcommand{\tess}{\textit{TESS}\xspace}
\newcommand{\gaia}{\textit{Gaia}\xspace}

\newcommand{\nkois}{235\xspace}
\newcommand{\nstars}{2658\xspace}

\received{October 4, 2024}
\accepted{January 7, 2025}

\shorttitle{Exoplanet Occurrence Rate with Stellar Age}
\shortauthors{Sayeed et al.}

\begin{document}

\title{Exoplanet Occurrence Rate with Age for FGK Stars in \textit{Kepler}}


\correspondingauthor{Maryum Sayeed}
\email{maryum.sayeed@columbia.edu}

\author[0000-0001-6180-8482]{Maryum Sayeed}
\affiliation{Department of Astronomy, Columbia University, 550 West 120th Street, New York, NY, USA}

\author[0000-0003-4540-5661]{Ruth Angus}
\affiliation{Department of Astrophysics, American Museum of Natural History, 200 Central Park West, Manhattan, NY, USA}
\affiliation{Center for Computational Astrophysics, Flatiron Institute, 162 5th Avenue, Manhattan, NY, USA}
\affiliation{Department of Astronomy, Columbia University, 550 West 120th Street, New York, NY, USA}

\author[0000-0002-2580-3614]{Travis A. Berger}
\affiliation{Space Telescope Science Institute, 3700 San Martin Drive, Baltimore, MD 21218, USA}

\author[0000-0003-4769-3273]{Yuxi(Lucy) Lu}
\affiliation{Department of Astrophysics, American Museum of Natural History, 200 Central Park West, Manhattan, NY, USA}
\affiliation{Department of Astronomy, The Ohio State University, Columbus, 140 W 18th Ave, OH 43210, USA}

\author[0000-0002-8035-4778]{Jessie L. Christiansen}
\affiliation{NASA Exoplanet Science Institute, IPAC, MS 100-22, Caltech, 1200 E. California Blvd, Pasadena, CA 91125}

\author[0000-0002-9328-5652]{Daniel Foreman-Mackey}
\affiliation{Center for Computational Astrophysics, Flatiron Institute, 162 5th Avenue, Manhattan, NY, USA}

\author[0000-0001-5082-6693]{Melissa K. Ness}
\affiliation{Research School of Astronomy \& Astrophysics, Australian National University, Canberra, ACT 2611, Australia}
\affiliation{Department of Astronomy, Columbia University, 550 West 120th Street, New York, NY, USA}

\begin{abstract}
We measure exoplanet occurrence rate as a function of \iso and \gyro ages using confirmed and candidate planets identified in Q1--17 DR25 \kepler data. We employ \kepler's pipeline detection efficiency to correct for the expected number of planets in each age bin. We examine the occurrence rates for planets with radii $0.2 \leq Rp \leq 20$ \earthrad and orbital periods $0.2 \leq P \leq 100$ days for FGK stars with ages between $1.5-8$ Gyr using the inverse detection efficiency method. We find no significant trend between occurrence rate and stellar ages; a slight, decreasing trend (within 1.5--2.5 $\sigma$) only emerges for low--mass and metal--rich stars that dominate our sample. We isolate the effects of mass and metallicity on the \orate trend with age, but find the results to be inconclusive due to weak trends and small sample size. Our results hint that the exoplanet occurrence rate may decrease over time due to dynamical instability from planet--planet scattering or planet ejection, but accurate ages and larger sample sizes are needed to resolve a clear relation between \orate and age.
\end{abstract}


\section{Introduction} \label{sec:intro}
NASA's \kepler mission revolutionized the field of exoplanets, and motivated population studies to understand the \orate and diversity of planets in the Galaxy \citep{kepler_borucki_2010, kepler_koch_2010}. The number and diversity of exoplanets have enabled studies across multiple planet parameters, such as planet size, orbital period, and mass (see \cite{Mulders_2018} and \cite{Zhu2021} for a review and references therein). Many have also considered the effects of stellar properties -- given the mutual relationship between planets and their host stars -- such as
metallicity \citep[e.g.,][]{FischerValenti, Petigura2018, Boley2024}, 
stellar mass (and effective temperature) \citep[e.g.,][]{Yang2020}, and binaries \citep[e.g.,][]{maxwell_2021}.
Others have focused planet population studies for a specific spectral type of host star, such as M--dwarfs \citep[e.g.,][]{Hardegree-Ullman2019, ment2023}, or stellar populations, such as thin and thick disk \citep[e.g.,][]{Zink2023}, and halo stars \citep[e.g.,][]{Boley2021}. NASA's \tess mission has already found hundreds of planets in its short tenure thus far \citep[][]{tess_ricker}, motivating recent planet \orate studies with \tess data \citep[e.g.,][]{Beleznay2022, Gan2023, Temmink_2023}. However, while these studies inform us about the occurrence of diverse planet types, they provide little clues into exoplanet system evolution over time. 

Investigation of planetary system evolution requires an understanding of planet demographics as a function of stellar ages. Quantifying the relationship between age and the \orate of planets allows us to constrain the frequency of planets lost due to dynamical instability, planet engulfment, or other processes, while also constraining the prevalence of free--floating planets. However, detailed studies of exoplanet \orate with age are rare given the absence of accurate ages for field stars in large numbers. Theoretical models of planet formation rely on the timescales at which planets evolve, which is difficult to constrain without confidence in age. Although stellar mass and metallicity can be used as age proxies, their effects must be isolated to probe age dependence. 

Two popular methods to infer stellar ages are isochrone fitting and \gyro. Isochrone fitting works well for stars in clusters, as well as for older and more massive stars \citep[e.g.,][]{Soderblom2010}. This method enables straightforward derivation of stellar parameters with input observables, such as parallax, photometry, and stellar metallicity, combined with a vast grid of stellar models. Alternatively, \gyro relies on the relationship between stellar rotation and age \citep{Skumanich1972}; stars lose angular momentum as they progress through their lifetime due to the interaction between magnetic fields and stellar winds \citep[e.g.,][]{Kawaler1988, Barnes2003, vanSaders2016, Metcalfe2019, Hall2021, Saunders2024} which can be leveraged to derive ages for thousands of stars. However, \gyro relies strongly on empirical calibration. Gyro--kinematic ages \citep[][]{Angus2020, lucy2021}, which combine the vertical velocity dispersion of a star with other measurable stellar properties (such as the rotation period, effective temperature, $G$ magnitude, and Rossby number) to derive stellar ages, have recently been used to calibrate a consistent \gyro model to infer \gyro ages for single field stars \citep{lucy2024}. 

To date, there have been a handful of studies of exoplanet \orate and stellar ages \citep[e.g.,][]{Hamer2019, Trevor_2021, Swastik_2023, Christiansen2023, Vach2024, Yang2023} and Fernandes et al. (submitted). However, the majority of these studies have investigated young planets in clusters and associations; for instance, \cite{Christiansen2023} studied the occurrence rate of hot, sub--Neptunes in Praesepe with an age of $< 800$ Myr, while \cite{Vach2024} focused systems with ages below 200 Myr. 

Motivated by the recent breakthrough in \gaia's parallax measurements, as well as the recent availability of stellar ages for a large sample of \kepler stars, we explore the relationship between exoplanet \orate and stellar age up to 8 Gyr. Although previous studies have used a variety of methods to measure the \orate, such as approximated Bayesian computation \citep[e.g.,][]{dfm2014, Bryson2020, Kunimoto_2020, KunimotoBryson2020, Shabram_2020}, we employ the inverse \deteff method, which relies on the completeness of the planet search algorithm in \kepler \citep{burke2015, Christiansen2020}. Recent analyses with K2 and \tess have focused on short--period planets; our work extends to longer periods which may resolve the evolution of planet radius valley with time. Understanding the correlation between stellar age and exoplanet \orate can inform our target selection of future planet--search surveys as well as provide insight into planet formation.

\begin{figure*}[t!]
    \centering
    \includegraphics[width=\linewidth]{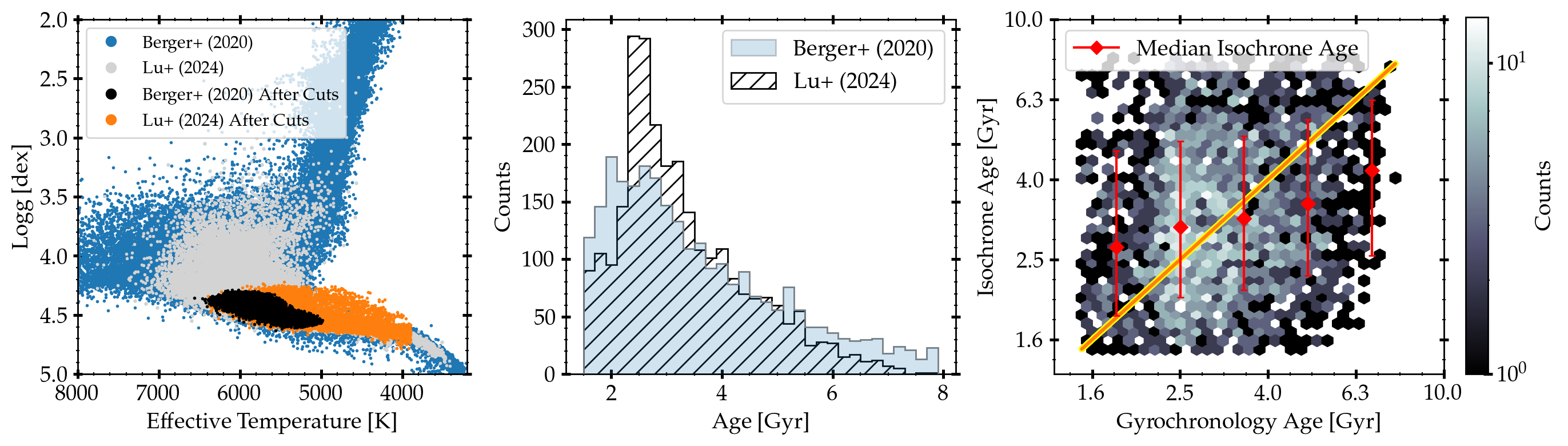}
    \caption{\textit{Left:} Stellar sample from B20 and L24 on an Hertzprung--Russell diagram before and after quality cuts. \textit{Middle:} Age distribution of isochrone and \gyro ages for \nstars stars in both sets. \textit{Right:} Comparison of \gyro and isochrone ages with an RMS of 1.79 Gyr. The red points show the median \iso and \gyro age within age bins where error bars show the 16th and 84th percentile of \iso age. The yellow line indicates perfect agreement between the two age samples.} There is a significant deviation between both ages for overlapping stars; stars with \gyro ages are preferentially younger than those with isochrone ages, and cover a larger parameter space than isochrone ages.
    \label{fig:properties}
\end{figure*}

\section{Sample Selection}
We construct a sample of stars with available and reliable \iso ages from \cite{Berger_2020} (hereon referred to as B20) and \gyro ages from \cite{lucy2024} (hereon referred to as L24) for confirmed or candidate \kepler planets. We briefly describe age determination method for both the \iso and \gyro age samples below, and refer the reader to B20 and L24 for more a more detailed overview. 
\begin{enumerate}[i)]
    \item Isochrone ages were derived from a custom--interpolated MIST \citep{Paxton_2011,Paxton_2013,Paxton_2015,Dotter_2016,Choi_2016} grid of \app7 million models with ages between $0.1-20$ Gyr and $-2.0-0.5$ dex in [Fe/H]. The code \texttt{isoclassify} was used to derive stellar parameters -- including age -- using input observables: photometry (SDSS $g$ and 2MASS $K_s$), \gaia DR2 parallaxes, red giant evolutionary flags, and spectroscopic metallicities. 
    \item Gyrochronology ages were derived using a newly calibrated \gyro relation that is most reliable for single field dwarf stars between $1.5-14$ Gyr of solar metallicity. Extra rotation period measurements were first obtained using the Zwicky Transient Facility \citep[ZTF;][]{ztfdata, ztftime}, and the relation is then calibrated using gyro--kinematic ages \citep[][]{Angus2020, lucy2021} and known cluster members \citep[][]{Curtis2020, Dungee2022} with a 2D Gaussian process using \texttt{tinygp} \citep{tinygp}. Testing on reproducing asteroseismic ages \citep[][]{SilvaAguirre2017} shows a median absolute deviation of 1.35 Gyr, and testing age predictions for wide binary pairs \citep[][]{Gruner2023, El-Badry2021} agree within 0.83 Gyr.
\end{enumerate}


Quality cuts in both the B20 and L24 samples were performed to ensure a high fidelity sample. In the B20 sample, we select stars with \gaia re--normalized unit weight error (RUWE) $\leq$ 1.2 (to exclude potential binaries and contaminated photometry), terminal age main--sequence (TAMS) less than 20 Gyr, and model goodness--of--fit above 99\%. This reduces the B20 sample from 186,301 to 119,518 stars. We use stars with available spectroscopic metallicity to ensure a good age constraint (as indicated in B20), reducing the sample to 40,213 stars, and require that error on TAMS is below 50\% to reduce biases towards older stars. For L24 sample, we select stars with RUWE $\leq$ 1.2 to remove potential binaries.

We further reduce both stellar samples to a specific region of the HR diagram, selecting stars with \teff between $3900-7300$ K as the definition of FGK stars from \cite{Kunimoto_2020} and stellar radii (fit with isochrones from B20) below 1.15 \solrad. Furthermore, to ensure high confidence in planet detection, we require a data span of at least two years, and duty cycle above 60\%. We also limit the sample to stars found in both B20 and L24 in order to derive a relationship between \orate and age consistent with both samples. Finally, we restrict the samples to ages between $1.5-8$ Gyr, given the inaccuracy of gyrochronology ages below 1.5 Gyr. Of the 7362 stars with \iso ages and 17,192 stars with \gyro ages, \nstars stars are found in both B20 and L24. 

Figure \ref{fig:properties} shows the B20 and L24 samples on an Hertzprung--Russell diagram before and after quality cuts (left panel), as well as the age distribution for \nstars overlapping stars (middle and right panels). While \iso ages span the entire age range, there are less older stars with \gyro ages. The right panel of Figure \ref{fig:properties} shows the large disagreement in \iso and \gyro age for the same stars, demonstrating the difficulty in age derivation for dwarf field stars. The red points show the median \iso and \gyro age within five equal age bins (in log space). The error bars indicate the 16th and 84th percentile of \iso ages. The median absolute deviation of the residual is 1.05 Gyr. It is important to note that neither the \iso nor the \gyro ages are the `ground truth' in this work. Therefore, the exaggerated level of scatter in the right panel is expected since we are plotting an uncertain measurement against another uncertain measurement, rather than comparing an uncertain quantity to a confident measurement. Furthermore, both the \iso and \gyro methodologies used to produce ages here were calibrated and compared favorably to open clusters and asteroseismic constraints (see \cite{Berger_2020} and \cite{lucy2024} for more details). As such, while the large disagreement between the \iso and \gyro ages suggests randomness in ages used in this work, they are calibrated on more confident age measurements.

To create our planet sample, we select confirmed and candidate \kepler planets in Q$1-17$ in Data Release 25 on the NASA Exoplanet Archive, restricted to planets with orbital periods between $0.2-100$ days and planet radii between $0.2-20$ \earthrad for our \nstars hosts; \nkois planets satisfy these conditions.  



\section{Methods}\label{sec:methods}


\subsection{Detection Efficiency}

For a single target, we can calculate the \deteff, or \kepler pipeline completeness, over a grid of planet radii and orbital periods. The completeness of the \kepler survey relies on three individual probabilities: the detection probability (\Pdet), geometric probability (\Pgeom), and the window function (\Pwin). The completeness can therefore be defined as, 
\begin{equation}
    P_\mathrm{comp}=P_\mathrm{geom} \times P_\mathrm{det} \times P_\mathrm{win}
    \label{eq:pcomp}
\end{equation}
We briefly describe the analytic forms of each probability below, and direct the reader to \cite{burke2015} for a more complete discussion.

The geometric probability, \Pgeom, is given by,
\begin{equation}
    P_\mathrm{geom} = \frac{1}{a/R_\star}\frac{1}{1-e^2}.
    \label{eq:pgeom}
\end{equation}

and requires as input the semi--major axis of the orbit $a$, the stellar radius $R_\star$, and eccentricity $e$ \citep{kipping2014}. We assume a circular orbit for our analysis, and discuss the shortcomings of this later.

Given broad--band red noise, as is the case with \kepler lightcurves, the Transiting Planet Search algorithm (TPS) considers the Multiple--Event Statistic (MES) to measure the strength of a potential transit signal. In the presence of a signal, the MES is distributed as a Guassian with unit variance, and an average proportional to the signal--to--noise ratio of the transit signal. The MES is defined as the average of the transit signal strength over multiple transit events,
\begin{equation}
    \mathrm{MES}=\sqrt{N_\mathrm{tr}}\frac{\Delta}{\sigma_\mathrm{cdpp}}, \; N_\mathrm{tr}=\frac{T_\mathrm{obs}}{P_\mathrm{orb}}\times f_\mathrm{duty}
    \label{eq:mes}
\end{equation}

where $N_\mathrm{tr}$ is the expected number of transit events, $T_\mathrm{obs}$ is the transit baseline, and $f_\mathrm{duty}$ is the observing duty cycle. The expected transit signal depth, $\Delta$, is 
\begin{equation}
    \begin{split}
    & \Delta = 0.84\Delta_\mathrm{max}, \\
    & \Delta_\mathrm{max}  = k^2(c+sk), \\
    & k = Rp/R_\star
    \end{split}
    \label{eq:depth}
\end{equation}

where $c$ and $s$ are fit by a linear relationship that vary with the limb darkening profile of stellar intensity. For G dwarfs, \cite{burke2015} determine the best--fit values to be $c=1.0874$ and $s=1.0187$. In Equation \ref{eq:mes}, $\sigma_\mathrm{cdpp}$ refers to the Combined Differential Photometric Precision \citep[CDPP,][]{kepler_koch_2010, Christiansen_2012}, as the time varying noise in the light curve, averaged over the relevant transit duration. For a given transit duration, $\tau_\mathrm{dur}$, we interpolate within a grid of 14\footnote{1.5, 2.0, 2.5, 3.0, 3.5, 4.5, 5.0, 6.0, 7.5, 9.0, 10.5, 12.0, 12.5, 15.0 in hours.} robCDPP values, to estimate the noise ($\sigma_\mathrm{cdpp}$) for that duration, where transit duration takes as input \Porb, $a$, $e$, and $R_\star$, and is defined as,
\begin{equation}
    \tau_\mathrm{dur}=4\Bigg(\frac{P_\mathrm{orb}}{1\; \mathrm{day}}\Bigg) \Bigg(\frac{R_\star}{a}\Bigg)\sqrt{1-e^2} \; \mathrm{hr.}
\end{equation}

The \Pdet, which is only a function of MES, is defined as the fraction of transit signals present in the data that are recovered by TPS, given by,
\begin{equation}
    P_\mathrm{det}(\mathrm{MES}) = \frac{1}{2}+\frac{1}{2} \mathrm{erf}\Bigg[\frac{(\mathrm{MES-MES_{thresh}})}{\sqrt{2}}\Bigg]
    \label{eq:pdet}
\end{equation}

where MES$_\mathrm{thresh}=7.1$ \citep[][]{Jenkins_2002a, Jenkins_2002b}. However, relying on only MES results in a over 50\% false detection rate. The observed pipeline completeness is therefore suppressed by the theoretical expectation, which can be quantified by
\begin{equation}
    P_\mathrm{gamma}(x | a,b) = \frac{1}{b^a\Gamma(a)}\int_{0}^{x} t^{a-1}\exp^{-t/b}\;dt
    \label{eq:pgamma}
\end{equation}
where $x=\mathrm{MES}-4.1-(\mathrm{MES_{thresh}} - 7.1)$, $a=4.65$ and $b=0.98$. 

\begin{figure}[t!]
    \centering
    \includegraphics[width=\linewidth]{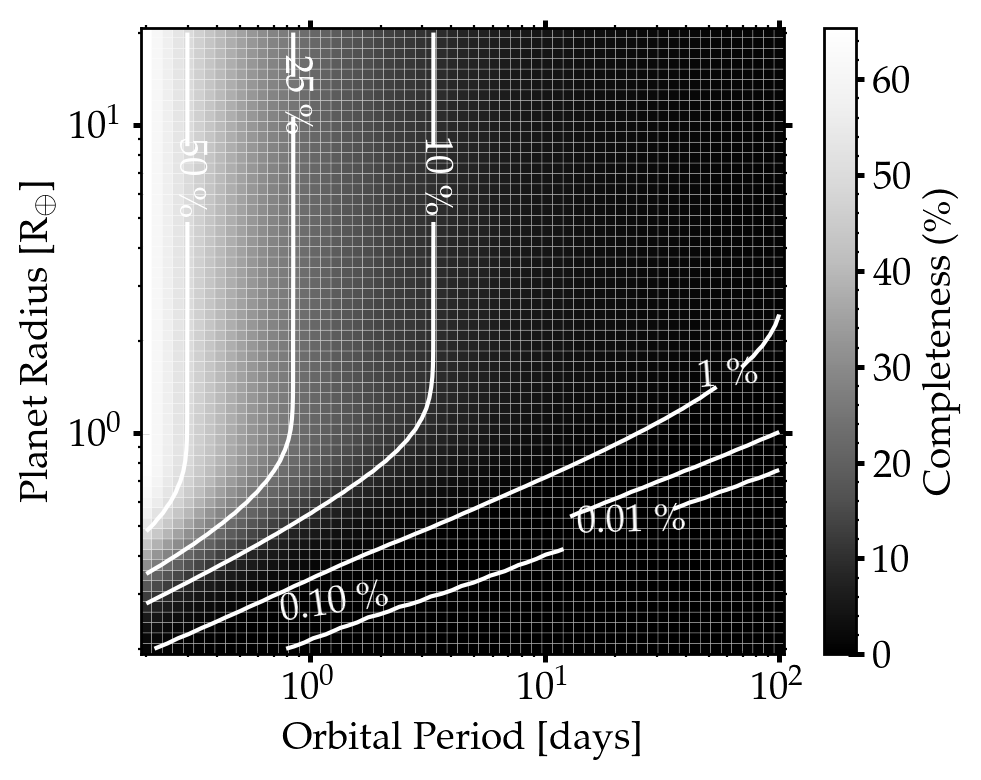}
    \caption{Integrated \deteff for the full sample of \nstars with available isochrone and \gyro ages stars over a grid of orbital period and planet radii. Contours and gridlines are shown for reference.}
    \label{fig:Q}
\end{figure}

\begin{deluxetable*}{c|cccc|cccc}[t!]
\tabletypesize{\footnotesize}
\tablecaption{Number of planets (observed and corrected) and stars, as well as the mean age error, in each age bin for sample of isochrone and gyrochronology ages used in this work. \label{tb:age_stats}}
\tablehead{\multicolumn{1}{c|}{} &  \multicolumn{4}{c|}{Isochrone Ages from \cite{Berger_2020}} &  \multicolumn{4}{c}{Gyrochronology Ages from \cite{lucy2024}} \\
\multicolumn{1}{c|}{Age [Gyr]} & \colhead{$\sigma_\textrm{age}$ [Gyr]} & \colhead{Observed $N_\textrm{Planets}$} & \colhead{Corrected $N_\textrm{Planets}$} & \multicolumn{1}{c|}{$N_\textrm{Stars}$} & \colhead{$\sigma_\textrm{age}$ [Gyr]} & \colhead{Observed $N_\textrm{Planets}$} & \colhead{Corrected $N_\textrm{Planets}$} & \colhead{$N_\textrm{Stars}$}}
\startdata
$1.500-2.096$ & 2.35 & 46 & 73 & 454 & 0.15 & 28 & 47 & 288 \\
$2.096-2.930$ & 3.09 & 55 & 114 & 720 & 0.07 & 87 & 155 & 974 \\
$2.930-4.095$ & 3.89 & 57 & 103 & 655 & 0.14 & 61 & 128 & 799 \\
$4.095-5.723$ & 4.61 & 61 & 84 & 534 & 0.27 & 50 & 74 & 475 \\
$5.723-8.000$ & 4.83 & 16 & 47 & 294 & 0.44 & 9 & 19 & 122 \\
\enddata
\end{deluxetable*}
\vspace*{-\baselineskip}

Lastly, the transit window function, \Pwin, accounts for the probability that the number of transits required for detection, $N_\mathrm{tr}\geq3$, occurs in the observational data. The analytic form for \Pwin can be approximated as a binomial,
\begin{equation}
    \begin{split}
    P_\mathrm{win,\geq3}=1 & - (1-f_\mathrm{duty})^M - Mf_\mathrm{duty}(1-f_\mathrm{duty})^{M-1} \\
    & - \frac{M(M-1)}{2}\;f^2_\mathrm{duty}(1-f_\mathrm{duty})^{M-2}
    \end{split}
    \label{eq:pwin}
\end{equation}
where $M=T_\mathrm{obs}/P_\mathrm{orb}$. \cite{burke2015} found a negligible difference in the pipeline completeness for shorter periods, \Porb $\leq 300$ days, when using the analytic window function (Equation \ref{eq:pwin}) and a more accurate numerical pipeline completeness model. 

Therefore, taking each of the \nstars stars as input, we calculate the detection probability over a grid of planet properties. We construct a grid with 61 logarithmically--spaced orbital period bins between 0.2 and 100 days, and 60 logarithmically spaced planet radius bins between 0.2 and 20 \earthrad, where the choice of radius bins is similar to limits chosen by \cite{Mulders_2018}. Although many planet demographic studies select planets at orbital periods within 10 days, we chose 100 days as the upper limit as 10 days was too restrictive for our sample \citep{Fulton2017}. This grid choice ensures that smaller periods are sampled more than larger periods over the wide range of periods, which accounts for the fact that close--in planets are more common in the data. Similarly, even though larger planets are more common in the data than smaller planets, the large number of bins in the narrow range of \rp ensures that \rp parameter space is also well sampled. 

Figure \ref{fig:Q} shows the combined completeness for the sample of \nstars stars in our sample. The completeness is constant for all periods for larger planets, above \app 3 \earthrad. Our detection efficiency contours are similar to those of \cite{Mulders_2018} and \cite{Dattilo_2023}, where the latter found a similar result for larger planets. As expected, we find a higher efficiency for larger and closer planets since those are easier to detect. 


\subsection{Inverse Detection Efficiency Method}
The \orate, $\Gamma$, using the inverse detection efficiency method is described by the following equation,
\begin{equation}
    \Gamma_i = \frac{n_{i,\mathrm{planets}}}{N_{i,\mathrm{stars}}}\frac{1}{\Bar{Q}_i}
    \label{eq:gamma}
\end{equation}
where $i$ denotes a given bin in age, $Q_i$ is the \deteff, and $\Gamma_i$ is the \orate. For each bin, we calculate the observed rate of planets per star $n_\mathrm{planets}/N_\mathrm{stars}$, and correct for this observed rate by dividing by the mean \deteff $Q$ in the respective bin. In Equation \ref{eq:gamma}, $n_\mathrm{planets}$ are the uncorrected number of planets. To summarize, we calculate \orate in three steps:
\begin{enumerate}[a)]
    \setlength\itemsep{0em}
    \setlength\parskip{0em}
    \item for all stars, $j$, in a given age bin, we find the mean $Q$ over the grid (e.g., $Q_1$, $Q_2$, ... $Q_j$), then calculate the average \deteff of those stars, $\Bar{Q}_i$ in a given bin,
    \item we calculate the rate of observed number of planets and stars in each bin, $\frac{n_{i,\mathrm{planets}}}{N_{i,\mathrm{stars}}}$,
    \item we correct for the observed rate of planets per star by the \deteff, $\Bar{Q}_i$
\end{enumerate}

The distribution of $Q$ values for a given star over the \rp$-P$ grid (3660 values) produced a right skewed histogram, instead of a normal distribution. We compared the mean, median, and mode of stars in each age bin, and found that the distribution of $Q$ values across age bins did not deviate; however, the range of each of mean, median, and mode varied significantly with ranges between $0.10-0.18$, $0.04-0.08$, and $0.40-0.85$, respectively. We experimented with different summary statistics, but the relative shape of the final distribution of \orate did not deviate; only the absolute scale of \orate varied based on the statistic used (ie. when using the median $Q$, the \orate was higher than when using the mode $Q$). Therefore, we chose to calculate the mean $Q$ since the values were not on either extremes.


\begin{figure*}[t!]
    \centering
    \includegraphics[width=\linewidth]{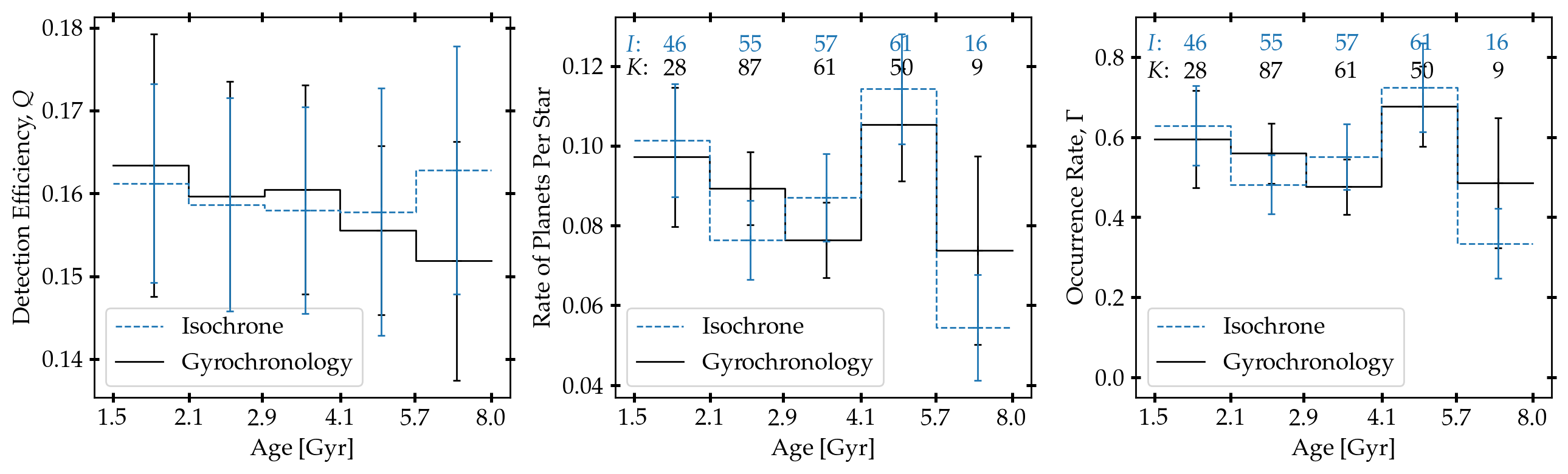}
    \caption{Planet occurrence rate as a function of isochrone (blue, dashed line) and gyrochronology (black, solid line) ages. \textit{Left:} detection efficiency, $Q$, per age bin where the error bars are the standard deviation of $Q$ in each bin. \textit{Middle:} number of planets per star where error bars indicate Binomial errors (Equation \ref{eq:sigmaR}). \textit{Right:} \orate of planets per age bin where the error bars indicate propagated error (Equation \ref{eq:gamma_err}) from detection efficiency and rate of planets. The text in the middle and right panels indicate the number of observed planets in each bin for stars with \iso ages (blue) and \gyro ages (black); see Table \ref{tb:age_stats} for number of stars in each bin. We find no significant trend in exoplanet \orate as a function of age for both samples.}
    \label{fig:results}
\end{figure*}

\subsection{Uncertainties}\label{sec:error}

We describe the uncertainty in rate of observed planets per star by a Binomial distribution where in a given bin, $k$ is the number of successes (or planets discovered), $n$ is the number of trials (or stars found), and $p$ is the success rate (or rate of planets per star). The standard error in the rate of planets per star is given by the following,
\begin{equation}
    \sigma_R = \sqrt{\frac{p(1-p)}{n}}
    \label{eq:sigmaR}
\end{equation}

The error on $\Bar{Q}_i$ is the standard deviation of the $\Bar{Q}_i$ distribution. As a result, the relative uncertainty on the \orate $\Gamma$ can be found via error propagation,
\begin{equation}
    \sigma_\Gamma = \Gamma \times \sqrt{\Big(\frac{\sigma_R}{R}\Big)^2 + \Big(\frac{\sigma_Q}{Q}\Big)^2 }
    \label{eq:gamma_err}
\end{equation}
where $R$ and $\sigma_R$ denote the rate of observed planets and its error, respectively.



\section{Results}\label{sec:results}

\subsection{Trends with Age}

We now apply our method to derive planet \orate as a function of age given \nkois (candidate or confirmed) \kepler planets and \nstars stars with \iso and \gyro ages, binned in five bins equally spaced logarithmically. Table \ref{tb:age_stats} shows the bin limits and average age error, and the number of planets and stars in each bin for both \iso and \gyro samples from B20 and L24, \resp. 

Figure \ref{fig:results} shows the result of our analysis: from left to right, we show the \deteff, rate of planets per star, and \orate as a function of \iso age in blue and \gyro age in black. Although the \deteff as a function of \iso age increases but decreases as a function of \gyro ages (as seen in the first panel), the \deteff from both samples is within $1-$sigma of each other. The middle panel of Figure \ref{fig:results} shows the rate of planets per star. Table \ref{tb:age_stats} lists the number of planets and stars in each age bin for both age samples.

The last panel in Figure \ref{fig:results} shows the \orate over time when using both \iso and \gyro ages. To quantify the trend shown in Figure \ref{fig:results}, we use weighted least squares linear regression \citep{lr_book} to calculate the slope of the distribution using the following equation,

\begin{equation}
    \hat{\beta} = (X^T W X)^{-1} X^T W y
    \label{eq:lr}
\end{equation}

where $\hat{\beta}$ is the vector of estimated coefficients, $X$ is the matrix of independent variables (or age in this work), $W$ is a diagonal matrix of weights, and $y$ is the dependent variable (or $\Gamma$ in this work). We use the inverse of the $\sigma_\Gamma$ to calculate weight, rather than the inverse of variance of $\sigma_\Gamma$. We also calculate the \pval where lower \pvals correspond to more statistically significant trends, in particular if \pval $<0.05$. In Figure \ref{fig:results}, the slope of \orate with age is $-0.03\pm0.04$ and  $-0.01\pm0.02$ for \iso and \gyro age samples, \resp. The \pvals for both distributions are large ($0.45$ and $0.73$, \resp) which suggests that the trends are not significant when accounting for uncertainties on the \orate. Table \ref{tb:slope} lists the slope with uncertainties and \pvals in Figure \ref{fig:results}.

However, derived stellar ages are also dependent on mass and metallicity, which could bias the overall \orate. For instance, given that older stars are rarely metal--rich, and exoplanet \orate decreases with decreasing metallicity for giant planets \citep[e.g.,][]{FischerValenti}, an apparent decrease in \orate with age could be due to decreasing metallicity and not age. The opposite occurs for stellar mass, where the \orate of exoplanets decreases with increasing stellar mass \citep[e.g.,][]{Johnson2010, Yang2020}. Since more massive stars tend to be younger and to have fewer planets, \orate--age trends caused by mass would resemble exoplanet \orate increasing with stellar age. In order to determine the causality of the age trend, we isolate the effects of mass and metallicity by deriving the \orate in bins of age, \iso mass from \cite{Berger_2020}, and spectroscopic metallicity from \cite{Berger_2020}. 

\begin{figure}[t!]
\centering
\includegraphics[width=\linewidth]{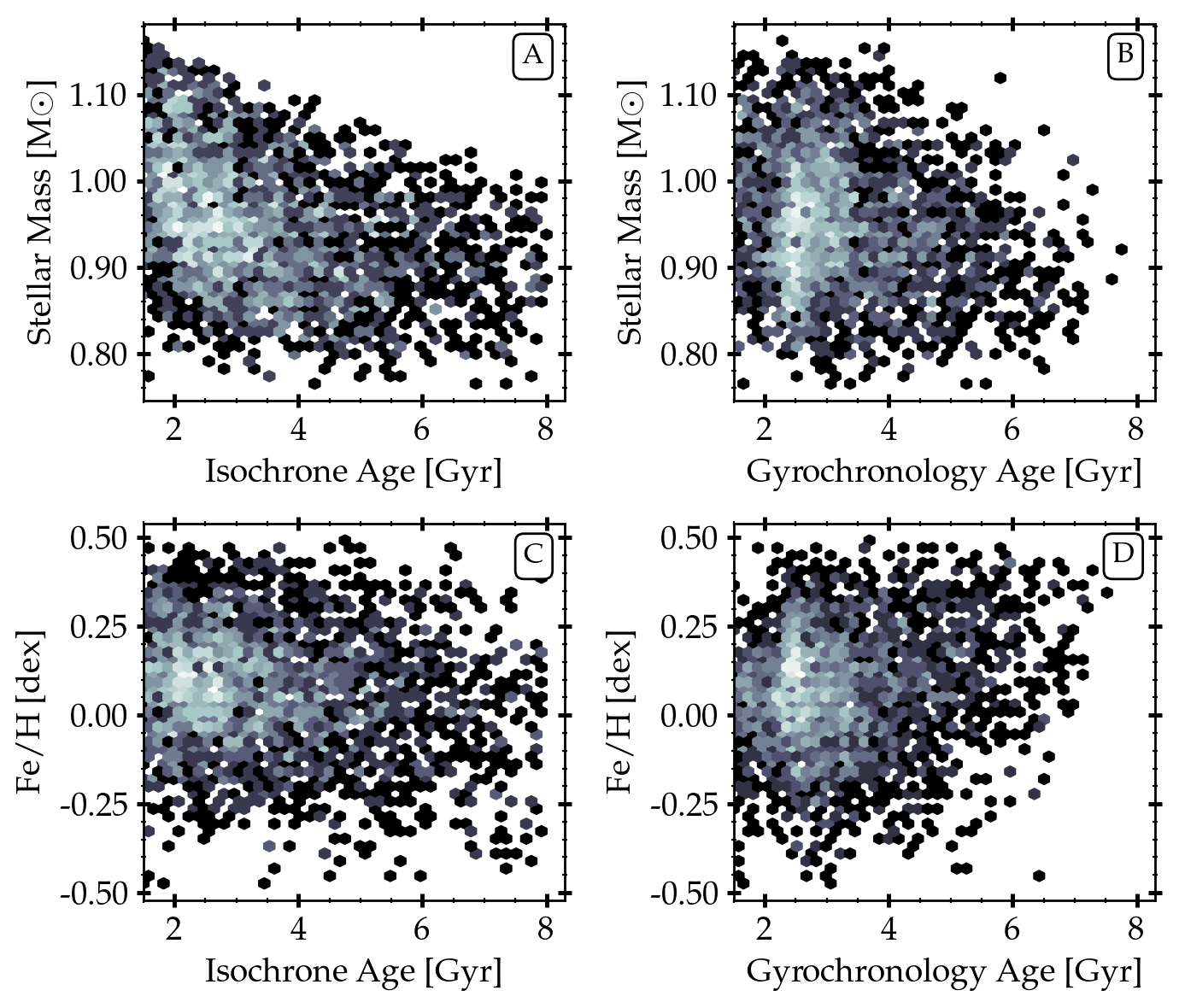}
\caption{Distribution of stellar \iso mass (top) and spectroscopic metallicity (bottom) with \iso (left) and \gyro (right) ages where the colour indicates the number of data points in each bin, with equal number of bins across the four distributions. These distributions illustrate that the two samples of \iso and \gyro ages span a different parameters space in mass and metallicity despite sharing the same stars. It also shows the relationship between mass and metallicity with age which could bias the trend between \orate and age.}
\label{fig:params_with_age}
\end{figure}

In Figure \ref{fig:params_with_age}, we show the distribution of stellar mass (top) and metallicity (bottom) as a function of \iso (left) and \gyro (right) ages. The number density of stars in the \iso sample peaks at around solar mass and metallicity at \app2 Gyr while for the \gyro sample, the peak is at a range of masses and metallicity. The peak at \app 2.5 Gyr is understood as the age for a typical star in the \kepler field as also found by other studies \citep[e.g.,][]{Pinsonneault2018}. Furthermore, the pile--up at \app2.5 Gyr is expected as stars go through stalled spin--down around that age \citep{Curtis2020} which means that stars older than 2.5 Gyr appear to be younger. The number density of stars with stellar mass decreases with age for both \iso and \gyro samples as expected (Panels A \& B in Figure \ref{fig:params_with_age}). Furthermore, Figure \ref{fig:params_with_age}D shows a slight trend where metallicity increases for older ages. This could be due to a detection bias where metal--poor stars -- which have shallower convective zones and weaker dynamo -- are less likely to produce star spots which enable a rotation period measurement \citep[e.g.,][]{See2021}. Metal--rich stars however would have stronger dynamo and are more likely to produce star spots. Since \gyro ages are dependent on a rotation period measurement, the lack of metal--poor older stars in Figure \ref{fig:params_with_age}D could be caused by a detection bias in \cite{lucy2021}. Figure \ref{fig:params_with_age} shows the existing trends with age that could bias the \orate with age trends. Despite sharing the same stars, the masses and metallicities span a different regime for the two samples of \iso and \gyro ages.

\begin{figure*}[t!]
    \centering
    \includegraphics[width=\linewidth]{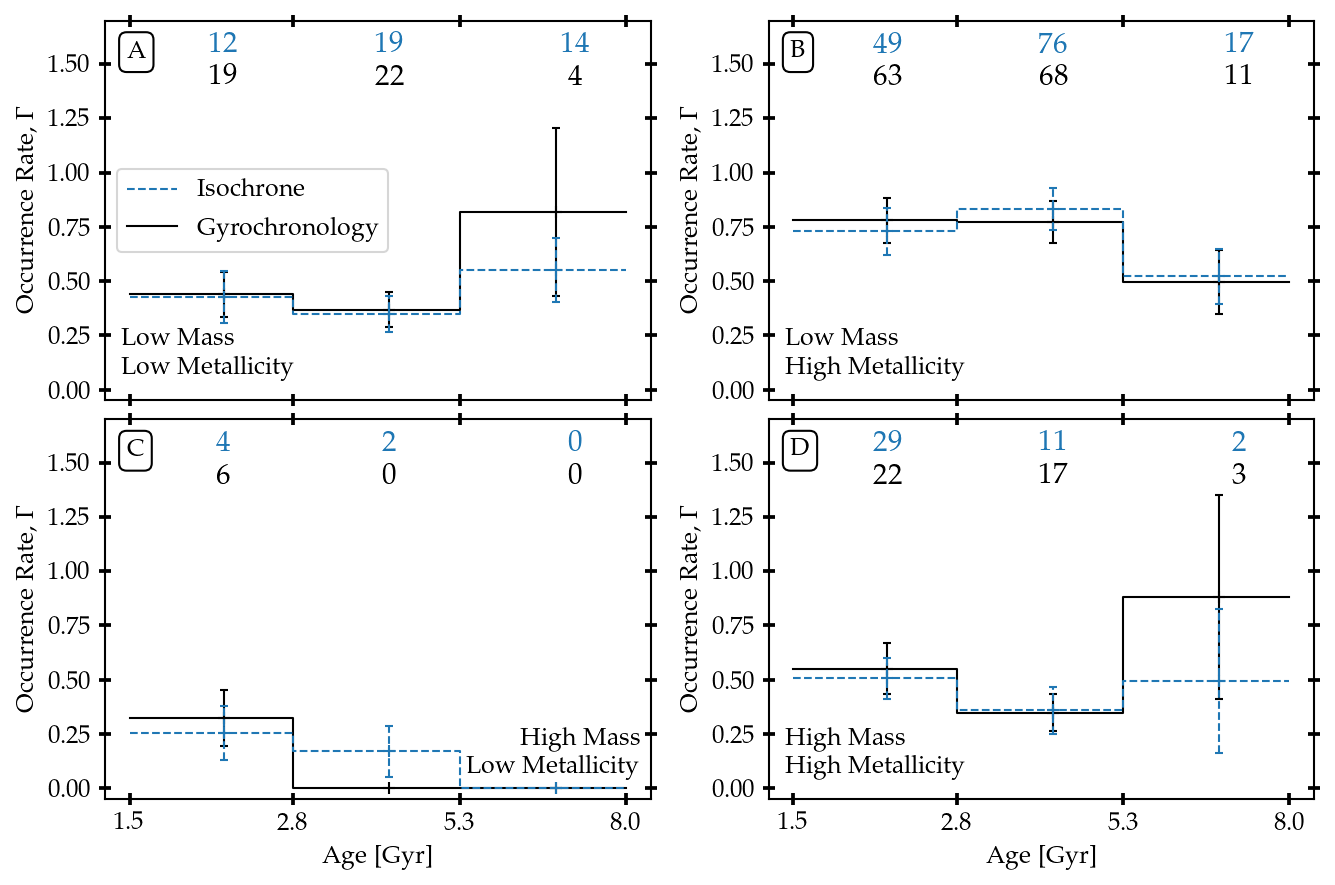}
    \caption{Planet \orate as a function of \iso (blue, dashed line) and \gyro (black, solid line) ages in groups of stellar mass and metallicity. The text indicates the number of observed planets in each bin. The mass increases from top to bottom, and the metallicity increases from left to right where the low and high mass samples refer to mass ranges of $0.8-1.0$ \solmass and $1.0-1.2$ \solmass, \resp, and low to high metallicity refers to metallicity ranges of $-0.5-0.0$ dex and  $0.0-0.5$ dex, \resp. Based on the slope and \pval (see Table \ref{tb:slope}), there are no significant trends in bins of mass, metallicity, and age.}
    \label{fig:mass_feh}
\end{figure*}


We incorporate more bins in mass and metallicity in Figure \ref{fig:mass_feh}, where we derive \orate in four bins to isolate their effects on the \orate for both \iso (blue, dashed lines) and \gyro (solid, black lines) ages; the stellar mass increases with rows (top to bottom), while the stellar metallicity increases with columns (left to right). The two bins in mass are $0.8-1.0$ \solmass and $1.0-1.2$ \solmass, while the two bins in metallicity are $-0.5-0.0$ dex and  $0.0-0.5$ dex. Therefore, we evaluate the \orate as a function of age (\iso and \gyro) in four bins: (i) low--mass, metal--poor (Panel A), (ii) low--mass, metal--rich (Panel B), (iii) high--mass, metal--poor (Panel C), and (iv) high--mass, metal--rich (Panel D).

Our results are inconclusive for three of the four bins in mass and metallicity: Panels A, C \& D. Firstly, for low--mass, metal--poor (Panel A) and high--mass, metal--rich (Panel D) samples, the uncertainties on the \orate for older stars ($5.3-8.0$ Gyr) are significantly large enough to make the trend indistinguishable for both samples. This is consistent with their slope and \pval in Table \ref{tb:slope}; for instance, the slopes using \iso and \gyro samples are $0.027\pm0.025$ (\pval $=0.48$) and $0.077\pm 0.041$ (\pval $=0.31$) in Figure \ref{fig:mass_feh}A, and $0.003\pm0.029$ (\pval $=0.93$) and $0.071\pm0.64$ (\pval $=0.47$) in Figure \ref{fig:mass_feh}D. Secondly, for high--mass, metal--poor stars (Panel C), the \orate is zero for the last bin in \iso age and the latter two bins in \gyro age given there are no planets in this range of mass, metallicity, and age; as a consequence, no slope or \pval can be derived for the two distributions in Panel C. The remaining bin for which the trend is significant in slope is for the \gyro age sample in Figure \ref{fig:mass_feh}B with a slope of $-0.044\pm0.036$, where the \orate with age decreases for low--mass, metal--rich stars. However, the \pval is 0.205 which suggests that the trend is not statistically significant.


Age trends are visible in samples of restricted metallicity ranges, which suggests that the apparent decrease in planet \orate as a function of age may indeed be causally driven by age and not metallicity. If metallicity alone were responsible for making older stars appear to have fewer planets, one would expect to see a difference in the occurrence rates of different metallicity hosts, particularly in the oldest bin. The oldest bins in the top row panels of Figure \ref{fig:mass_feh} are consistent within uncertainties, indicating that metallicity may not be a confounder. Furthermore, we also acknowledge that two broad bins in mass and metallicity may not be sufficient to reveal mass and metallicity trends. With more precise mass and metallicity measurements and a larger number of bins, future studies may be able to disentangle the effects of age, mass, and metallicity more definitively.

\subsection{Assumptions \& Caveats}\label{sec:assumption}

A drawback of the inverse \deteff method is that we do not incorporate errors on planet and stellar properties when calculating detection efficiency, $Q$. Errors on $P$ are negligible (average of 0.001\%), but on \rp are on average 19\%, with some errors above 50\%. The fractional error on \rp are greatest between 1--5 \earthrad. The average uncertainty on stellar radius, metallicity, and mass from \cite{Berger_2020} are \app2.5\%, \app3.2\%, and \app5.0\% for \nstars in our sample, respectively, and therefore do not contribute significantly to the uncertainty on \orate.

Consequently, the quantifiable uncertainty on \orate, $\Gamma$, is driven by the relative uncertainties on the rate of planets per star and detection efficiency. The uncertainties on the former are between $5-12$\% for the majority of the bins, while the error on $Q$ is between $5-10$\% for all bins. The error on \orate, $\Gamma$, is largely driven by the error on the rate of planets, and is approximately $10-13\%$ for \iso ages and $8-16\%$ for \gyro ages. While the error on age does not contribute to the quantifiable uncertainty on \orate, $\Gamma$, the large uncertainty impacts which age bin each star is assigned, and therefore has a major effect on the rate of planets per star. 


Age uncertainties are on average 56\% for the \iso sample which can significantly impact which bin the star is found in. While \gaia has enabled age determination for billions of stars with precise magnitudes and astrometry, these ages are not necessary meaningful or accurate because of the degeneracies associated with determining a stellar age from a photometric color, flux, parallax, metallicity, and a set of isochrone models. See B20 for more details on age determinations and the systematics/caveats therein. As is demonstrated by \cite{Tayar_2022}, differences between the exact set of isochrone models used can change derived ages by \app$20$\%, and this is generally for main--sequence and subgiant stars. In addition, depending on where a star is located within the HR diagram, the usefulness of isochrone ages can vary widely. For low--mass stars that do not evolve appreciably within the age of the Universe, isochrone ages are unconstrained, while for high--mass stars, the models are not as effective at reproducing observed stellar properties, in addition to the lower number of planets found around those stars. 

Gyrochronology ages are calibrated on an age--velocity relation that is suitable for metal--rich stars, [Fe/H] $> -0.2$ dex \citep{YuLiu2018}, which would impact older stars; however, only 11\% of our sample is metal--poor (329/\nstars) and does not contribute significantly to our results. Moreover, the model does not take into account that metallicity can affect rotation period of a star \citep{Amard2020} which could introduce large uncertainties up to 2 Gyr in inferring \gyro ages in this case \citep[e.g.,][]{Claytor2020, lucy2024}. Binary interactions or planet engulfment could also spin up stars and introduce extra uncertainty.

Furthermore, in our analysis we assume an eccentricity of zero which has direct impact on the geometric probability, \Pgeom in Equation \ref{eq:pgeom}. However, \cite{kipping2014} showed that when a non--negligible eccentricity is included, \Pgeom is enhanced by \app 10\%. While this would scale our final \orate, it does not affect the overall trend with stellar age. Furthermore, \cite{Kipping2013} found that shorter period planets are often on circular orbits. Therefore, while the zero eccentricity assumption has negligible effect on short--period planets, it is possible that incorporating non--zero eccentricity for long period planets could increase the final detection probability, to be larger than a few percent level. Again, this would scale our final results but have negligible effects on the overall trend.

\subsection{Comparison with Literature} \label{sec:lit_comparison}


Not many studies have considered the effects of age on \orate across a wide range of ages, given the difficulty in measuring stellar ages. Recently, \cite{Zink2023} and \cite{Bashi2022} compared planet hosts in the young, thin disk and old, thick disk; they found that the \orate of close--in super Earths (\rp $=1-2$ \earthrad and $P=1-100$ days) are higher in the thin disk.
\cite{Haywood2008, Haywood2009} find that larger planets (R $\gtrsim 2$ \earthrad) are more common around older stars than smaller planets. Similarly, \cite{Chen2022} find Sub--Neptunes and giant planets are more common around thick disk stars, which hosts older stellar population. \cite{Hamer2019} found hot Jupiter host stars are preferentially younger as compared to a sample of field stars without hot Jupiters in \gaia DR2. 

Using MIST--MESA models to estimate age for a sample of 2611 exoplanet hosting stars in \gaia DR3, \cite{Swastik_2023} find that stars hosting giant planets are younger than those hosting smaller planets. More recently, \cite{Yang2023} found a decreasing trend between \orate and \gyro ages for their sample of \kepler planets (\rp $= 0.5-6$ \earthrad, $P=1-100$ days). Similar to us, they also re--evaluated the trend after removing the effects of other stellar parameters, such as mass and metallicity, and found that while the trend does not disappear, the difference in apparent \orate between younger and older planets becomes smaller.

While others have also looked at \orate of planets with age, the sample selection differs greatly from ours. For instance, \cite{Sandoval2021} found stars older than 3 Gyr are more likely to host super--Earths (\app 1.5 \earthrad) while younger stars (below 3 Gyr) are more likely to host sub--Neptunes (\app 2.5 \earthrad). 
Similarly, \cite{Berger_2020} also looked at the ratio of super--Earths to sub--Neptunes as a function of \iso ages, and found that the ratio of super--Earths to sub--Neptunes is higher among older stars than for younger stars. However, although we both use \iso ages, their conclusion was based on comparing stars in two age bins: younger and older than 1 Gyr. While our sample also consists of ages above 1 Gyr, we do not have any hosts below this age threshold.



\section{Planetary System Evolution} \label{sec:discussion}

Based on the results in Figure \ref{fig:mass_feh}, we find tentative evidence for a decreasing trend between \orate of planets with age driven by low--mass, metal--rich stars. Our results suggest that either planetary systems form similarly and then evolve over time, or planetary systems that form 8 Gyr ago are different from planetary systems that form 1.5 Gyr ago. Although planets can be lost through varying mechanisms after the initial phase of planet formation, there is low probability of planet addition (through formation or stellar flybys) after the circumstellar disk has cleared, which occurs in the first few million years \citep[e.g.,][]{Richert2018}. However, planets can be lost via planet engulfment, planet--planet scattering, or planet ejection throughout the system's lifetime. We briefly describe each mechanism below, and its expected impact on the \orate of exoplanets.

\begin{enumerate}
    \item \textbf{Planet Engulfment:} \cite{Liu2024} suggest that planet engulfment occurs for 8\% of late--F to G--dwarfs, while \cite{Spina2021} suggest that this probability is closer to $20-35$\%. However, planet engulfment is more likely during the giant phase, when the star expands and engulfs close--in planets, less probable for our sample of FGK stars. \cite{Oetjens2020} showed that more massive, metal--poor stars with smaller convective envelopes evolve faster on the main--sequence, which in turn favours a faster planet engulfment. Furthermore, planet engulfment timescales are shorter for more massive host stars. 
    \item \textbf{Planet--Planet Scattering:} Planet--planet scattering is more likely but would occur during the early stages of planet formation in the presence of protoplanetary disk, planetesimals, gas, dust and other planet forming materials; in fact, N--body simulations show that scattering events occur within the first 100 Myr of system formation \citep{Izidoro2021, Bitsch2023}. 
    \cite{PuWu2015} show that \kepler systems with one or two planets were descendants of closely packed multi--planet systems ($3-5$ planets) that have undergone dynamical instability. Furthermore, \cite{Yang2023} found a decreasing trend between planet multiplicity and \gyro age which suggests that the probability of planet--planet scattering also decreases with age.
    \item \textbf{Planet Ejection:} Planet ejection can occur through stellar flybys or dynamical instability. It differs from planet--planet scattering in that this mechanism removes a planet from the system, where as planet--planet scattering changes the inclination and order of the planets in a system. Stellar flybys are fairly likely given that most stars are born in clusters \citep{Wang_2023}; the multiplicity fraction is initially very high and decreases over time, and for FGK star remains $\gtrsim 50$\% \citep{Cuello2023}. In fact, \cite{Boley2021} find that $20-40$\% of field stars should have experienced at least one encounter (within 300 AU), and \cite{Malmberg2011} find that 78\% of stars with masses $0.8-1.2$ \solmass experience at least one fly--by. Simulations show that for solar--type stars ($0.8-1.2$ \solmass), the probability of encounter decreases with age of the cluster, from 50\% at 1.5 Myr to less than 5\% at 5 Myr \citep{Pfalzner2013}. 

\end{enumerate}

\section{Conclusion}
We derived the \orate of \kepler exoplanets with available \iso and \gyro ages from \cite{Berger_2020} and \cite{lucy2024} between $1.5-8$ Gyr. Using only stars with both \iso and \gyro ages (\nstars stars), we find no significant trend with \orate and stellar ages. We tested the effect of mass and metallicity on age trends, and found a slight, decreasing trend ($< 2\sigma$) for low--mass, metal--rich stars. We urge caution in over--interpreting our results given the large uncertainties in both \iso and \gyro ages, as well as our small sample size (\nkois planets and \nstars stars). We attempted to disentangle the effects of mass and metallicity (see Figure \ref{fig:mass_feh}), but evaluating the \orate along three axes -- mass, metallicity, and age -- significantly reduced our sample size, especially for older stars ($5.3-8$ Gyr). 

Accurate ages for planet hosts are needed to effectively derive the relationship between planet \orate and age. Although asteroseismology is the only technique that can determine a star’s age with uncertainty as low as \app10\% \citep[e.g.,][]{Bellinger2019}, these are not available for large sample of \kepler stars. Age proxies such as lithium abundance can also be used to study \orate given the anti--correlation between lithium and age \citep[e.g.,][]{Skumanich1972}, but is out of the scope of this paper. Our sample is restricted to FGK stars with both \iso and \gyro age indicators where the performance of our chosen age methodologies is not as accurate as in other regions of the HR diagram. 

Fortunately, the primary science goal of upcoming surveys is to detect exoplanets and oscillations in solar--like stars which will significantly increase our confidence in planet \orate studies with age. The primary goal of the upcoming PLAnetary Transits and Oscillations of stars ({\small PLATO}) mission is the search for terrestrial planets around solar--like stars \citep{Plato2014, Plato2016}. In fact, \cite{Boettner2024b} predict that {\small PLATO} will detect 13,000 planets across the three major Galactic environments, the young, thin disk, and the older thick disk and halo, with the majority found in the young, thin disk. Furthermore, asteroseismology is a core science component of {\small PLATO}, which is expected to measure oscillation frequencies for 15,000 dwarf and subgiant stars with $V<11$. With precise ages provided by asteroseismology, {\small PLATO} will prove instrumental in planet demographic studies as a function of stellar age. Similarly, the Nancy Grace Roman Space Telescope \citep[\textit{Roman},][]{Spergel2015, Akeson2019} has a dedicated Galactic Time--Domain Bulge Survey that is expected to yield \app$60,000-200,000$ of transiting planets towards the galactic bulge \citep{Wilson2023}. Given that \textit{Roman} is expected to yield \app$10^6$ asteroseismic detections in the center of the galaxy \citep{Huber2023}, we can expect thousands of exoplanets with precise ages that would enable further \orate studies. Accurate stellar ages are crucial in understanding the evolution of planetary systems over time. Upcoming missions dedicated to the discovery and characterization of exoplanets and their host stars will reveal valuable insight into planet formation and planetary system evolution with help from planet population studies.

\section{Acknowledgments}
\begin{acknowledgments}
We are thankful to our anonymous referee for helpful comments which improved this manuscript. MS thanks the LSSTC Data Science Fellowship Program, which is funded by LSSTC, NSF Cybertraining Grant \#1829740, the Brinson Foundation, and the Moore Foundation; her participation in the program has benefited this work. 
\end{acknowledgments}

\software{astropy \citep{Astropy}, matplotlib \citep{matplotlib}, SciPy \citep{Scipy}, statsmodels \citep{statsmodels} }

\begin{deluxetable*}{c|cc|rccc}[t!]
\tablecaption{Slope and $p-$value for each distribution shown in Figures \ref{fig:results} and \ref{fig:mass_feh} where the slope was calculated using Equation \ref{eq:lr}. Statistically significant results usually have a $p-$value below 0.05. \label{tb:slope}}
\tablehead{Ages & Mass Range & [Fe/H] Range & Slope & $\sigma_\textrm{Slope}$ & $p-$value & Figure \\ 
& [\solmass] & [dex] & Gyr$^{-1}$ & Gyr$^{-1}$ & & }
\startdata
Isochrone & -- & -- &  $-0.033$ & $0.038$ & $0.451$ & \ref{fig:results} \\
Gyrochronology & -- & -- &  $-0.009$ & $0.023$ & $0.731$ & \ref{fig:results} \\ \hline
Isochrone & Low & Poor &  $0.027$ & $0.025$ & $0.475$ & \ref{fig:mass_feh}A \\
Isochrone & Low & Rich &  $-0.044$ & $0.036$ & $0.433$ & \ref{fig:mass_feh}B \\
Isochrone & High & Poor &  & & & \ref{fig:mass_feh}C \\
Isochrone & High & Rich & $0.003$ & $0.029$ & $0.932$ & \ref{fig:mass_feh}D \\
Gyrochronology & Low & Poor & $0.077$ & $0.041$ & $0.313$ & \ref{fig:mass_feh}A \\
Gyrochronology & Low & Rich &  $-0.055$ & $0.018$ & $0.205$ & \ref{fig:mass_feh}B \\
Gyrochronology & High & Poor & & & & \ref{fig:mass_feh}C \\
Gyrochronology & High & Rich &  $0.071$ & $0.064$ & $0.467$ & \ref{fig:mass_feh}D \\ \hline
\enddata
\tablecomments{`Low' and `high' mass ranges refer to masses $0.8-1.0$ \solmass and $1.0-1.2$ \solmass, respectively. `Poor' and `rich' metallicity ranges refer to $-0.5-0.0$ dex and $0.0-0.5$ dex, respectively.}
\end{deluxetable*}
\vspace*{-\baselineskip}

\bibliography{references}{}
\bibliographystyle{aasjournal}


\end{document}